\documentclass{iaus}
\usepackage{graphicx}

\title[water maser
] %% give here short title %%
{22GHz water maser survey of Xinjiang Astronomical Observatory \\
}

\author[Jian-jun. Zhou \& Jarken. Esimbek \& Gang Wu]   %% give here short author list %%
{Jian-jun. Zhou$^{1,2}$
 \ Jarken. Esimbek$^{1,2}$
 \and Gang. Wu$^{1,2}$}

\affiliation{$^1$Xinjiang Astronomical Observatory, CAS \\
150 Science 1 street  Urumqi, Xinjiang 830011, China \\ email: {\tt zhoujj@xao.ac.cn} \\[\affilskip]
$^2$Key Laboratory of Radio Astronomy, CAS, \\
150 Science 1 street  Urumqi, Xinjiang 830011, China \\}
\pubyear{2012}
\volume{287}  %% insert here IAU Symposium No.
\pagerange{119--126}
 \jname{Cosmic masers-from OH to H$_{\circ}$}
\editors{Roy. Booth, Liz. Hmphries \& W.H.T. Vlemmings, eds.}
\begin{document}

\maketitle

\begin{abstract}
Water masers are good tracers of high-mass star-forming regions.
Water maser VLBI observations provide a good probe to study
high-mass star formation and the galactic structure. We plan to make
a blind survey toward the northern Galactic plane in future years
using 25m radio telescope of Xinjiang Astronomical Observatory. We
will select some water maser sources discovered in the survey and
make high resolution observations and study the gas kinematics close
to the high-mass protostar. 
\end{abstract}

\firstsection % if your document starts with a section,
\section{Introduction}

High-mass star forming regions are usually at far distances, heavy
obscuration makes it difficult to observe them. The water masers are
good probes of physical conditions and dynamics of the star forming
regions. Maser VLBI observations are the unique mean by which one
can explore the gas kinematics close (within tens or hundreds of AU)
to the forming high-mass protostar (\cite[Moscadelli et al.
2011]{Moscadelli_etal11}). Measure trigonometric parallaxes and
proper motions of water masers found in high-mass star-forming
regions by VLBI reference method can provide very accurate distance
of them. Combining positions, distances, proper motions and radial
velocities yields complete 3-dimensional kinematic information of
the Galaxy (\cite[Xu et al. 2006]{Xu_etal06}; \cite[Reid et al.
2009]{Reid_etal09}). Water masers are very rich in the Galaxy, they
are reliable tracers of high-mass star-forming regions
(\cite[Caswell et al. 2011]{Caswell_etal11}). Therefore, it is
valuable to discover more water masers associated with high-mass
star-forming regions.

Earlier water maser searches have chiefly been made to targeted
sources, many masers may not be discovered. There are only a few
unbiased water maser surveys (\cite[Breen \etal\
2007]{Breen_etal07}; \cite[Walsh et al. 2008]{Walsh_etal08};
\cite[Caswell \& Breen 2010]{CaswellBreen10}). Recently, one much
larger blind survey toward 100 square degree of southern Galactic
plane has been completed successfully (\cite[Walsh et al.
2011]{Walsh_etal11}). However, no large blind water maser survey has
been done toward northern Galactic plane. We will make a blind
survey toward 90 square degree of the northern galactic plane using
our 25m radio telescope. We hope to discover a large sample of water
masers and high-mass star-forming regions at earlier stages, and
study high-mass star formation and galactic structure.

\section{25m radio telescope of Xinjiang Astronomical Observatory}
Nanshan 25m radio telescope of Xinjiang Astronomical Observatory was
built in 1992 as a station for the Chinese very long baseline
interferometry network. It is located at Nanshan mountains west of
Urumqi city at an altitude of 2080m. Its front-end receiver system
includes several receivers working at 18, 13, 6, 3.6 and 1.3cm. At
1.3 cm, one dual-polarization cryogenic receiver has been installed
on the telescope recently, the noise temperature of the receiver is
better than 20K. When weather is good, the system temperature is
better than 50K. We built a molecular spectrum observing system in
1997. One digital filter bank (DFB) system is employed as the
spectrometer, it is is capable of processing up to 1 GHz of
bandwidth with 8192 channels. Our telescope now can observe several
molecules such as OH, H$_{2}$O, NH$_{3}$, H$_{2}$CO and
H$_{110\alpha}$.

\section{Our plan}

We will make a large scale blind survey toward Northern Galactic
plane. For that most water masers concentrated in the region along
the galactic plane ($|b|<0.5^{\circ}$). We plan to survey 90 square
degrees of the northern galactic plane, it covers the region between
l=30$^{\circ}$ and l=120$^{\circ}$,and $|b|<0.5^{\circ}$. In order
to complete the project in reasonable time, scan observation mode
(on the fly) will be used in our observation, and final sensitivity
of the survey is about 1.4Jy.

On the other hand, many surveys at millimeter, submillimeter,
infrared wavelengths discovered a large sample of possible
star-forming regions, e.g. Bolocam, Planck, Glimpse and MIPS. These
sources provide us good candidates for searching water masers. We
also can select some sources and make targeted survey.

\begin{acknowledgements}
This work was funded by The National Natural Science Foundation of
China under Grant $10778703$, China Ministry of Science and
Technology under State Key Development Program for Basic Research
(2012CB821800) and The National Natural Science Foundation of China
under Grant $10873025$.
\end{acknowledgements}

\end{document}